\documentclass[
,12pt]{article}
\usepackage[affil-it]{authblk}
\usepackage[english]{babel}
\usepackage{blindtext}
\usepackage{enumitem}
\usepackage{indentfirst}

\providecommand{\keywords}[1]{\textbf{\textit{Keywords:}} #1}

\usepackage{graphicx}
\usepackage{amsmath}
\usepackage{amsthm}
\usepackage{amssymb}
\usepackage{amsfonts}
\usepackage[hidelinks]{hyperref}

\usepackage{booktabs}

\usepackage{longtable}
\usepackage{cite}
\usepackage{mathrsfs}

\usepackage[title,titletoc,toc]{appendix}

\newtheorem{case}{Case}

\theoremstyle{definition} 

\usepackage{chngcntr}
\counterwithin{subcase}{case}
\counterwithin{exmp}{section}
\counterwithin{rem}{section}
\counterwithin{red}{section}
\counterwithin{prof}{section}

\usepackage[margin=1in,left=1in,right=1in]{geometry}
\usepackage{subcaption}
\usepackage{float}
\usepackage{placeins}
\allowdisplaybreaks
\usepackage[final]{changes}
\usepackage{appendix}
\usepackage{cleveref}
\usepackage{longtable}
\usepackage{rotating}
\usepackage{array}
\usepackage{booktabs}

\title{Lie symmetry classification and group invariant solutions of generalized radial heat equation with nonlinear reaction source}

\author[1]{Manjit Singh\thanks{corresponding author: manjitcsir@gmail.com}}
\author[2]{Radhika}

\affil[1,2]{%
    Yadavindra Department of Sciences, Punjabi University Guru Kashi Campus, Talwandi Sabo--151302, Punjab, India.}
\begin{document}
\maketitle
\begin{abstract}
This work presents a Lie symmetry classification of a generalized nonlinear heat equation with a reaction source term in radial geometry. The model involves three arbitrary constitutive functions that represent thermal capacity, thermal conductivity, and nonlinear heat generation or absorption. Using the classical Lie invariance criterion, the determining equations for point symmetries are derived and simplified through suitable transformations involving the ratios of the constitutive functions. The classification identifies several admissible subclasses for which the principal symmetry algebra is extended, including power-law and logarithmic branches associated with special values of the radial parameter. For these cases, the admitted Lie algebras, commutator structures, and optimal systems of one-dimensional subalgebras are obtained. The corresponding similarity reductions are constructed, reducing the governing partial differential equation to nonlinear ordinary differential equations. Some exact group-invariant solutions are also derived for special parameter choices. The results show that the inclusion of the nonlinear source term significantly enriches the symmetry structure compared with the source-free radial heat equation.

\end{abstract}

\section{Introduction}
Lie symmetry analysis has become one of the most powerful analytical tools
for the investigation of nonlinear partial differential equations arising
in mathematical physics, continuum mechanics, fluid dynamics, heat
transfer, and reaction--diffusion phenomena. Since the pioneering work of
Sophus Lie, continuous transformation groups have provided a systematic
framework for studying differential equations through their intrinsic
symmetry properties \cite{mansfield2,haydonbook,anco,NUCCI,olverbook,bruzon2001symmetry,bira2015exact,bluman2010applications,ames1992symmetry,pandey2008symmetry,pandey2009symmetry,faucher1993symmetry}. In particular, Lie symmetries enable the reduction of
partial differential equations to ordinary differential equations,
the construction of invariant solutions, the derivation of conservation laws.

For nonlinear evolution equations, an important aspect of Lie theory is the
group classification problem. Instead of analyzing a single equation with
fixed coefficients, it is more important to examine a class of equations containing
arbitrary constitutive functions or parameters and seeks to determine all
special forms for which the equations admit symmetry extensions beyond the
principal Lie algebra. Such classifications play a crucial role in the
systematic organization of nonlinear models because the admitted symmetry
structure is closely related to the existence of exact solutions,
similarity reductions  and conservation properties.

In diffusion and heat-transfer models, constitutive functions usually describe real physical properties, such as diffusivity, thermal capacity, conductivity, or nonlinear reaction sources. For this reason, classifying these constitutive relations is not just a mathematical exercise; it can also reveal which physical mechanisms are consistent with richer symmetry structures.
At the same time, arbitrary constitutive functions make the determining equations much harder to analyze. The infinitesimals get linked with unknown nonlinear coefficient functions, leading to complicated functional-differential relations.

The problem becomes even more complex in radial geometries. In these cases, the geometry itself adds extra terms that can create singular or exceptional cases for certain values of the radial parameter. These effects can change the admitted Lie algebra, give rise to resonant symmetry structures, and produce similarity reductions with very different behavior.
The situation is further complicated when nonlinear reaction sources are included. These sources create additional links between the symmetry generators and the constitutive functions, making the classification problem much harder than in the source-free case.

With these ideas in mind, this work focuses on the Lie symmetry classification of generalized radial heat equations with nonlinear reaction sources. The determining equations are obtained using the classical Lie invariance criterion and then studied systematically through suitable constitutive transformations and compatibility conditions.
The analysis identifies admissible subclasses that allow nontrivial extensions of symmetry. For these subclasses, the corresponding Lie algebras are constructed, and optimal systems of one-dimensional subalgebras are obtained. Finally, the admitted symmetries are used to derive similarity reductions and group-invariant solutions of the governing equation.

Motivated by the generalized nonlinear heat equation studied by Nauryz's \cite{nauryz2026lie}
for media with varying cross-section geometry, we consider an extended
radial heat-transfer model with nonlinear reaction effects. In particular,
Nauryz investigated the source-free equation
\begin{equation}
 C(u)u_t=
 \frac{1}{z^{\nu}}
 \left(z^{\nu}K(u)u_z\right)_z,
 \label{eq:nauryzmodel}
\end{equation}
where $C(u)$ and $K(u)$ denote temperature-dependent thermal coefficients,
and $\nu>0$ is a geometric parameter describing the effect of the varying
cross-section or radial geometry. Building on this framework, we study the
following reaction-diffusion extension:
\begin{align}
   \label{Rad:01}
   C(u)u_t
   =
   \frac{1}{z^{\nu}}
   \left(z^{\nu}K(u)u_z\right)_{z}
   +R(u).
\end{align}
This equation provides a flexible model for heat transfer and diffusion
processes in radial geometries. Here, $C(u)$ represents the thermal
capacity, $K(u)$ denotes the thermal conductivity, and $R(u)$ describes a
nonlinear reaction source. The function $R(u)$ allows the model to include
additional physical effects such as internal heat generation, absorption,
chemical reactions, or other source-driven processes.

Expanding the diffusion term in \eqref{Rad:01}, we obtain
\begin{align}
   C(u)u_t
   =
   K(u)u_{zz}
   +
   K'(u)u_z^2
   +
   \frac{\nu}{z}K(u)u_z
   +
   R(u).
   \label{eq:expandedpresent_model}
\end{align}
In this expanded form, the first two terms on the right-hand side describe
nonlinear diffusion and the contribution due to the temperature dependence
of the thermal conductivity. The term involving $\dfrac{\nu}{z}$ captures
the effect of radial geometry, where different values of $\nu$ correspond
to different spatial configurations.

The presence of arbitrary nonlinear functions $C(u)$, $K(u)$, and $R(u)$
makes the equation rich from both physical and mathematical points of
view. These functions allow the model to describe a wide range of
nonlinear diffusion and heat-transfer phenomena. At the same time, they
make the Lie symmetry classification problem more challenging, since the
admitted symmetries depend strongly on the specific forms of these
constitutive functions.

The main difference from the work of Nauryz is the inclusion of the
additional constitutive function $R(u)$. In the source-free case, the Lie
symmetry classification mainly depends on the relationship between
$C(u)$ and $K(u)$. In contrast, the present model requires the symmetry
generators to be compatible not only with $C(u)$ and $K(u)$, but also with
the nonlinear reaction source $R(u)$. This additional coupling makes the
determining equations more restrictive and substantially enriches the
classification problem. Consequently, the present study may lead to new
admissible subclasses, new symmetry extensions, and new similarity
reductions that do not appear in the source-free model.

For a broader discussion of the physical background, related nonlinear
heat-transfer models, and earlier applications of Lie symmetry methods to
generalized heat equations,  the references cited
in the work of Nauryz's can be seen.

\section{Lie Symmetry Formulation}

We consider the governing equation
\begin{equation}
C(u)u_t
=
K(u)u_{zz}
+
K'(u)u_z^2
+
\frac{\nu}{z}K(u)u_z
+
R(u).
\label{eq:governing_expanded}
\end{equation}
where \(C(u)\), \(K(u)\), and \(R(u)\) are smooth functions of \(u\), and \(\nu\) is a constant parameter.
Equivalently, we define
\begin{equation}
G(z,t,u,u_z,u_t,u_{zz})
=
C(u)u_t
-
K(u)u_{zz}
-
K'(u)u_z^2
-
\frac{\nu}{z}K(u)u_z
-
R(u),
\label{eq:G}
\end{equation}
so that the governing equation is simply
\begin{equation}
G(z,t,u,u_z,u_t,u_{zz})=0.
\label{eq:G_zero}
\end{equation}
Let a one-parameter local Lie group of point transformations act on the variables
\((z,t,u)\) as
\begin{equation}
\bar z=Z(z,t,u;\varepsilon),\qquad
\bar t=T(z,t,u;\varepsilon),\qquad
\bar u=U(z,t,u;\varepsilon),
\label{eq:group_transformations}
\end{equation}
where \(\varepsilon\) is the group parameter. The identity transformation is recovered at
\(\varepsilon=0\):
\begin{equation}
Z(z,t,u;0)=z,\qquad
T(z,t,u;0)=t,\qquad
U(z,t,u;0)=u.
\end{equation}
For small \(\varepsilon\), the transformations have the infinitesimal form
\begin{align}
\bar z
&=
z+\varepsilon \xi(z,t,u)+O(\varepsilon^2),
\\
\bar t
&=
t+\varepsilon \tau(z,t,u)+O(\varepsilon^2),
\\
\bar u
&=
u+\varepsilon \eta(z,t,u)+O(\varepsilon^2).
\end{align}
The corresponding infinitesimal generator is therefore
\begin{equation}
\mathcal{X}
=
\xi(z,t,u)\frac{\partial}{\partial z}
+
\tau(z,t,u)\frac{\partial}{\partial t}
+
\eta(z,t,u)\frac{\partial}{\partial u}.
\label{eq:infinitesimal_generator}
\end{equation}
Since the equation \eqref{eq:G} contains derivatives up to second order, the second prolongation of $\mathcal{X}$ is required. It is given by
\begin{equation}
\mathcal{X}^{(2)}
=
\mathcal{X}
+
\eta^z\frac{\partial}{\partial u_z}
+
\eta^t\frac{\partial}{\partial u_t}
+
\eta^{zz}\frac{\partial}{\partial u_{zz}},
\label{eq:second_prolongation}
\end{equation}
where the relevant prolonged coefficients are
\begin{align}
\eta^z
&=
D_z(\eta)-u_zD_z(\xi)-u_tD_z(\tau),
\label{eq:eta_z}
\\
\eta^t
&=
D_t(\eta)-u_zD_t(\xi)-u_tD_t(\tau),
\label{eq:eta_t}
\\
\eta^{zz}
&=
D_z(\eta^z)-u_{zz}D_z(\xi)-u_{zt}D_z(\tau).
\label{eq:eta_zz}
\end{align}
Here \(D_z\) and \(D_t\) denote the total derivative operators with respect to \(z\) and \(t\), respectively.
The invariance condition for the differential equation is
\begin{equation}
\mathcal{X}^{(2)}G\big|_{G=0}=0.
\label{eq:invariance_condition}
\end{equation}
Substituting these expressions into \eqref{eq:invariance_condition}, we obtain
\begin{align}
0
&=
\xi\left(
\frac{\nu}{z^2}K(u)u_z
\right)
\nonumber\\
&\quad
+
\eta\left(
C'(u)u_t
-
K'(u)u_{zz}
-
K''(u)u_z^2
-
\frac{\nu}{z}K'(u)u_z
-
R'(u)
\right)
\nonumber\\
&\quad
+
\eta^z
\left(
-2K'(u)u_z
-
\frac{\nu}{z}K(u)
\right)
+
C(u)\eta^t
-
K(u)\eta^{zz},
\qquad G=0.
\label{eq:invariance_expanded}
\end{align}
Finally, on the solution manifold \(G=0\), we may use
\begin{equation}
u_t
=
\frac{1}{C(u)}
\left[
K(u)u_{zz}
+
K'(u)u_z^2
+
\frac{\nu}{z}K(u)u_z
+
R(u)
\right].
\label{eq:ut_replacement}
\end{equation}
Substitution of \eqref{eq:ut_replacement} into \eqref{eq:invariance_expanded}, followed by splitting with respect to the independent derivatives, yields the determining equations for the infinitesimals
\(\xi(z,t,u)\), \(\tau(z,t,u)\), and \(\eta(z,t,u)\).

\begin{subequations}
    \begin{align}
    &\label{DE01}\xi_{u}=\,0, \tau_{z}=\,0, \tau_{u}=\,0\\[1mm]
&\label{DE02}\frac{\eta\, C'\,K}{C}-\eta \,K'-\tau_{t}\, K+2\,\xi_z\, K=0,\\[1mm]
&\label{DE03}K'\eta_u+\tau_{t} K'-2K'\xi_z+\eta K''+\eta_{uu}K
-\frac{\eta \,C'\,K'}{C}=0,\\[1mm]
&\label{DE04}C\eta_t - K\eta_{zz} - \frac{\nu K}{z}\eta_z + R\eta_u - \tau_t R
+ \eta \left( \frac{C'}{C}R - R' \right)=0,\\[1mm]
&C\,\xi_t-K\,\xi_{zz}+2K\,\eta_{zu}
+2\eta_z \,K'
-\frac{\nu\,\eta\, C'\,K}{z\,C}
-\frac{\nu\,\xi\,  K}{z^{2}}\nonumber\\[1mm]
\label{DE05}&-\frac{\nu\,\xi_z\, K}{z}+\frac{\nu\,\eta K'}{z}
+\frac{\nu\,\tau_{t}\, K}{z}=0.
\end{align}
\end{subequations}

The determining equation \eqref{DE02} can be simplified as: 
\begin{align}
    \label{rad:2}&\tau_{t}-2\,\xi_{z}=\,\eta\,M(u),\quad F(u)=\,\left(\frac{C'}{C}-\frac{K'}{K}\right),\\
   \label{rad:3} &\eta=\,\frac{1}{F(u)}\,(\tau_{t}-2\,\xi_{z})=\,G(u)\, S(z,t),
\end{align}
and \eqref{rad:3} is the first simplification in the classification process. For convenience, we introduce the notation
\begin{align}
\label{rad:20}&A(u)=\frac{C(u)}{K(u)}, \quad B(u)=\frac{R(u)}{K(u)},\quad F(u)=\frac{C'(u)}{C(u)}-\frac{K'(u)}{K(u)},\\[1mm]
\label{rad:21}&G(u)=\frac1{F(u)},\quad S(z,t)=\tau_t-2\xi_z.
\end{align}

Substituting $\eta$ from \eqref{rad:3} into \eqref{DE03} gives
\begin{align}
    \label{rad:4}S\left[
G''
+\frac{K'}{K}G'
+\left(
\frac{K''}{K}
-\frac{C'}{C}\frac{K'}{K}
\right)G
\right]=0
\end{align}
 and \eqref{DE04} becomes
 \begin{align}
     CS_tG
-KS_{zz}G
-\frac{\nu K}{z}S_zG
+RSG'
-\tau_tR
+SG\left(
\frac{C'}{C}R-R'
\right)=0
 \end{align}
 divide by $K$ further simplifies the above equation into
 \begin{align}
     \label{rad:5}AS_tG
-S_{zz}G
-\frac{\nu}{z}S_zG
+BSG'
-\tau_tB
+SB
-SGB'=0.
 \end{align}
 The equations \eqref{rad:4} and \eqref{rad:5} will serve as pivots for classification. Finally, from \eqref{rad:3}, we have
 \begin{align*}
     \eta_z=S_zG=-2\xi_{zz}G, \eta_{zu}=S_zG'=-2\xi_{zz}G'
 \end{align*}
Substituting these expressions into \eqref{DE05} and simplifying gives
\begin{align}
   \label{rad:6} A(u)\xi_t
-\Lambda(u)\xi_{zz}
-\frac{\nu}{z^2}\xi
+\frac{\nu}{z}\xi_z=0, 
\end{align}
where $\Lambda(u)=1+4G'(u)+4\frac{K'}{K}G(u)$. The equation \eqref{rad:6} is the main classification equation between $\xi$ and the constitutive functions $C,K$. To obtain symmetries beyond the kernel, we need to impose the following;
\begin{align}
    \label{rad:9} \Lambda(u)=\,a\, A(u)+b,
\end{align}
 where $a$ and $b$ are constants. This will split \eqref{rad:6} into a pair of partial differential equations in $\xi$ as follows:
\begin{align}
    \label{rad:7}\xi_t
-a\,\xi_{zz}=0,\\
\label{rad:8}b\,\xi_{zz}
+\frac{\nu}{z^2}\xi
-\frac{\nu}{z}\xi_z=0.
\end{align}

The equation \eqref{rad:8} is the standard Euler's equation and can be solved followed by the restrictions from \eqref{rad:7} will give the following splitting cases:
\begin{case}\normalfont If
$a\neq0,\; b\neq\nu,\; \nu\neq0,\; \nu\neq3b$, then
\begin{align}
    \xi=c_1z,\\
    \eta=(\tau_t-2c_1)G(u).
\end{align}
Plugging these $\xi, \eta$ into \eqref{rad:5} gives
\begin{equation}
\tau_{tt}
+P(u)\tau_t
=2c_1\,Q(u).
\label{rad:10}
\end{equation}
where 
\begin{align}
    P(u)=\frac{\left(BG'-GB'\right)}{AG},\quad Q(u)=\frac{\left(BG'+B-GB'\right)}{AG}.
\end{align}
Now $\tau$ depends on $t$ only and $P, Q$ depend on $u$, therefore solution of equation \eqref{rad:10} splits into two cases:
\begin{enumerate}
    \item If $P(u), Q(u)$ are non constants, then we can invoke the condition
     $BG'+B-GB'=0$ to get 
     \begin{align}
         \label{rad:100}B=\,B_{0}\,AG,\quad R=\,B_{0}\,CG
     \end{align}
     along with \eqref{rad:4} and \eqref{rad:9} a substantial simplification can be achieved as followed:
     \begin{align}
         \label{rad:11} G'=\,\frac{b-1}{4}, K=\,K_{0}\,\mathrm{exp}\left(\frac{a\,A}{4}\right), C=\,K_{0}\,A\,\mathrm{exp}\left(\frac{a\,A}{4}\right)
     \end{align}
     the equation \eqref{rad:5} further simplifies to 
     \begin{equation}
AG\,\tau_{tt}
+\left(BG'-GB'\right)\tau_t
=0.
\label{rad:12}
\end{equation}
which is equivalent to 
\begin{equation}
\tau_{tt}
-B_{0}\,\tau_t
=0.
\label{rad:13}
\end{equation}
 Therefore, we get
 \begin{align*}
     \tau=c_0+c_2e^{B_0t},
 \end{align*}
  Finally the infinitesimals for this case will be:
  \begin{align}
      \begin{cases}
          \xi=\,c_{1}\,z\\
 \tau=c_0+c_2e^{B_0t}\\
          \eta=
\left(
B_0c_2e^{B_0t}-2c_1
\right)\left(\frac{b-1}{4}\,u+G_{0}\right).
      \end{cases}
  \end{align}
Once the infinitesimals \(\tau\), \(\xi\), and \(\eta\) are determined,
the corresponding Lie symmetry generators are obtained through the
standard infinitesimal operator
\[
\mathcal{X}=\tau\partial_t+\xi\partial_z+\eta\partial_u.
\]
Since the infinitesimals depend linearly on the arbitrary constants
\(c_0,c_1,c_2\), the admitted Lie algebra is obtained by decomposing
$\mathcal{X}$ with respect to these constants, thereby yielding a basis of
independent symmetry generators \cite{ovsi,bluman2010applications}. Therefore, the infinitesimals generators corresponding to these infinitesimals can be obtained as follows:
  
  \begin{align*}
  \mathfrak{g}^{(1)}=
  \begin{cases}
      \mathcal{X}_{1}^{(1)}&=\partial_{t},\\[1mm]
\mathcal{X}_{2}^{(1)}&=
z\partial_{z}
-2\left(
\frac{b-1}{4}u+G_{0}
\right)\partial_{u},\\[1mm]
\mathcal{X}_{3}^{(1)}&=
e^{B_{0}t}\partial_{t}
+B_{0}e^{B_{0}t}
\left(
\frac{b-1}{4}u+G_{0}
\right)\partial_{u}.
  \end{cases}
\end{align*}
The generator
\(\mathcal{X}_{1}^{(1)}\) represents temporal translations,
\(\mathcal{X}_{2}^{(1)}\) corresponds to a scaling (dilatation)
transformation involving the radial coordinate and the dependent variable,
while \(\mathcal{X}_{3}^{(1)}\) generates an exponential time-dependent
symmetry induced by the admissible reaction source. The geometric interpretation of the basis generators can be understood better by defining  Lie bracket  in usual way $[\mathcal{X}_{i}, \mathcal{X}_{j}]=\,\mathcal{X}_{i}\mathcal{X}_{j}-\mathcal{X}_{j}\mathcal{X}_{i}$, and  the resulting commutations are summarized in \autoref{tab:comm_case1}.
\begin{table}[htbp]
\centering
\caption{Commutator table for the Lie algebra
$\mathfrak{g}^{(1)}$.}
\label{tab:comm_case1}
\renewcommand{\arraystretch}{1.2}
\begin{tabular}{c|ccc}
\hline
$[\cdot,\cdot]$
& $\mathcal{X}_{1}^{(1)}$
& $\mathcal{X}_{2}^{(1)}$
& $\mathcal{X}_{3}^{(1)}$
\\
\hline
$\mathcal{X}_{1}^{(1)}$
& $0$
& $0$
& $B_0\mathcal{X}_{3}^{(1)}$
\\
$\mathcal{X}_{2}^{(1)}$
& $0$
& $0$
& $0$
\\
$\mathcal{X}_{3}^{(1)}$
& $-B_0\mathcal{X}_{3}^{(1)}$
& $0$
& $0$
\\
\hline
\end{tabular}
\end{table}

Here $\mathcal{X}_{2}$ spans the center of the algebra $\mathfrak g^{(1)}$ and it is isomorphic $A_{2.1}\oplus A_{1}$ (see reference \cite{patera1977subalgebras}).
 To construct an optimal system of one-dimensional subalgebras, we use the
adjoint action on general element of algebra, the procedure of which is well established in \cite{olverbook}. Consequently, an optimal system of one-dimensional subalgebras is
\begin{equation}
\left\{
\mathcal{X}^{(1)}_{2},
\quad
\mathcal{X}^{(1)}_{1}+\alpha \mathcal{X}^{(1)}_{2},
\quad
\mathcal{X}^{(1)}_{3}+\beta \mathcal{X}^{(1)}_{2}
\right\},
\qquad
\alpha,\beta\in\mathbb R.
\end{equation}

 \item  When $P(u)=p$ and $Q(u)=q$ are constants. Taking
\[
P(u)=\frac{BG'-GB'}{AG}=p,
\qquad
Q(u)=\frac{BG'+B-GB'}{AG}=q,
\]
does not lead to a genuinely richer symmetry class. Indeed, subtracting the two relations yields
\[
B=(q-p)AG.
\]
Substituting this expression for \(B\) into the first relation gives
\[
p=-(q-p),
\]
and consequently
\[
q=0.
\]
Therefore, the assumption that both \(P(u)\) and \(Q(u)\) are constants necessarily reduces to
\[
Q(u)=0,
\]
that is,
\[
BG'+B-GB'=0.
\]
Hence, the constant-ratio ans\"atze does not generate any new symmetry-enhancing subclass beyond the one already obtained from the simpler condition \(BG'+B-GB'=0\). 
 \end{enumerate}
 
\end{case}
Compared to the source-free equation examined in \cite{nauryz2026lie}, the addition of the reaction source term \(R(u)\) significantly changes the structure of the Lie classification problem. In addition to the thermal capacity
\(C(u)\) and conductivity \(K(u)\), one must now classify the reaction
function \(R(u)\). After introducing the reduced variables
\(A=C/K\), \(B=R/K\), and
\(G=\left(C'/C-K'/K\right)^{-1}\), the determining equations contain
additional contributions involving \(B\) and \(B'\). These terms couple the
infinitesimals directly to the reaction source and prevent the reduction of
the determining system to a single compatibility equation for \(A\) and \(G\). As a result, classification must be performed simultaneously with respect
to the triplet \((C,K,R)\), leading to new compatibility conditions and
subclasses that enhance the symmetry of reaction terms that are absent in the
source-free setting. 
\begin{case}\normalfont
If $a=0,\; b\neq\nu$, then

\begin{align}
    \xi=c_1z+c_2z^{\nu/b},\\
    \eta=
\left(
\tau_t
-2c_1
-2\frac{\nu}{b}c_2z^{\nu/b-1}
\right)G(u)
\end{align}
Substituting these expressions into \eqref{rad:5} and separating with
respect to the independent powers of \(z\), we obtain
\[
(m-1)(m+\nu-2)=0,
\qquad
BG'+B-GB'=0.
\]
Since \(b\neq\nu\), we have \(m\neq1\), and therefore
\[
m=2-\nu,
\qquad
b=\frac{\nu}{2-\nu},
\qquad
\nu\neq2.
\]
Moreover,
\[
BG'+B-GB'=0
\]
integrates to
\[
B=B_0AG \implies R=B_0CG.
\]
Under this condition, equation \eqref{rad:5} reduces to
\[
\tau_{tt}-B_0\tau_t=0,
\]
whose solution is
\[
\tau=
\begin{cases}
c_0+c_3e^{B_0t},
&
B_0\neq0,
\\[1mm]
c_0+c_3t,
&
B_0=0.
\end{cases}
\]
Furthermore, combining \eqref{rad:4} and \eqref{rad:9} yields
\[
G'=\frac{b-1}{4},
\qquad
G(u)=\frac{b-1}{4}u+G_0,
\]
and
\[
\frac{K'}{K}=\frac{aA}{4G}
=\frac a4 A',
\]
which gives
\[
K(u)=K_0e^{aA(u)/4},
\qquad
C(u)=K_0A(u)e^{aA(u)/4}.
\]
Consequently,
\begin{align}
\xi&=c_1z+c_2z^{\,2-\nu},
\nonumber\\[1mm]
\tau&=
\begin{cases}
c_0+c_3e^{B_0t},
&
B_0\neq0,
\\[1mm]
c_0+c_3t,
&
B_0=0,
\end{cases}
\nonumber\\[1mm]
\eta&=
\begin{cases}
\left(
B_0c_3e^{B_0t}
-2c_1
-2(2-\nu)c_2z^{\,1-\nu}
\right)\left(\frac{b-1}{4}u+G_0\right),
&
B_0\neq0,
\\[1mm]
\left(
c_3
-2c_1
-2(2-\nu)c_2z^{\,1-\nu}
\right)\left(\frac{b-1}{4}u+G_0\right),
&
B_0=0.
\end{cases}
\label{eq:branch-final}
\end{align}
Once the infinitesimals are obtained the admitted Lie algebra for \(B_0\neq0\) $$\mathfrak{g}^{(2a)}
=
\left\langle
\mathcal{X}_{1}^{(2a)},
\mathcal{X}_{2}^{(2a)},
\mathcal{X}_{3}^{(2a)},
\mathcal{X}_{4}^{(2a)}
\right\rangle$$
 is generated by
\begin{align}
\begin{cases}
\mathcal{X}_{1}^{(2a)}
=
\partial_t,
\\[2mm]
\mathcal{X}_{2}^{(2a)}
=
z\partial_z
-2H(u)\partial_u,
\\[2mm]
\mathcal{X}_{3}^{(2a)}
=
z^{\,2-\nu}\partial_z
-2(2-\nu)z^{\,1-\nu}H(u)\partial_u,
\\[2mm]
\mathcal{X}_{4}^{(2a)}
=
e^{B_0t}\partial_t
+B_0e^{B_0t}H(u)\partial_u.
\end{cases}
\label{alg:2a}
\end{align}
and for \(B_0=0\), the admitted Lie algebra
\[
\mathfrak{g}^{(2b)}
=
\left\langle
\mathcal{X}_{1}^{(2b)},
\mathcal{X}_{2}^{(2b)},
\mathcal{X}_{3}^{(2b)},
\mathcal{X}_{4}^{(2b)}
\right\rangle
\]
is generated by
\begin{align}
\begin{cases}
\mathcal{X}_{1}^{(2b)}
=
\partial_t,
\\[2mm]
\mathcal{X}_{2}^{(2b)}
=
z\partial_z
-2H(u)\partial_u,
\\[2mm]
\mathcal{X}_{3}^{(2b)}
=
z^{\,2-\nu}\partial_z
-2(2-\nu)z^{\,1-\nu}H(u)\partial_u,
\\[2mm]
\mathcal{X}_{4}^{(2b)}
=
t\partial_t
+H(u)\partial_u.
\end{cases}
\label{alg:2b}
\end{align}
 where following notation is used \[
H(u)=\frac{b-1}{4}u+G_0.
\]
The nonvanishing Lie commutators of the basis generators
of $\mathfrak{g}^{(2a)}$ and $\mathfrak{g}^{(2b)}$
are summarized in Tables~\ref{tab:comm_2a}
and~\ref{tab:comm_2b}, respectively.
These commutation relations are subsequently employed
to construct the corresponding optimal systems of
one-dimensional subalgebras.
\begin{table}[htbp]
\centering
\caption{Commutator table for the Lie algebra
$\mathfrak{g}^{(2a)}$.}
\label{tab:comm_2a}
\renewcommand{\arraystretch}{1.2}
\begin{tabular}{c|cccc}
\hline
$[\cdot,\cdot]$
& $\mathcal X_{1}^{(2a)}$
& $\mathcal X_{2}^{(2a)}$
& $\mathcal X_{3}^{(2a)}$
& $\mathcal X_{4}^{(2a)}$
\\
\hline
$\mathcal X_{1}^{(2a)}$
& $0$
& $0$
& $0$
& $B_0\mathcal X_{4}^{(2a)}$
\\
$\mathcal X_{2}^{(2a)}$
& $0$
& $0$
& $(1-\nu)\mathcal X_{3}^{(2a)}$
& $0$
\\
$\mathcal X_{3}^{(2a)}$
& $0$
& $-(1-\nu)\mathcal X_{3}^{(2a)}$
& $0$
& $0$
\\
$\mathcal X_{4}^{(2a)}$
& $-B_0\mathcal X_{4}^{(2a)}$
& $0$
& $0$
& $0$
\\
\hline
\end{tabular}
\end{table}
\begin{table}[htbp]
\centering
\caption{Commutator table for the Lie algebra
$\mathfrak{g}^{(2b)}$.}
\label{tab:comm_2b}
\renewcommand{\arraystretch}{1.2}
\begin{tabular}{c|cccc}
\hline
$[\cdot,\cdot]$
& $\mathcal X_{1}^{(2b)}$
& $\mathcal X_{2}^{(2b)}$
& $\mathcal X_{3}^{(2b)}$
& $\mathcal X_{4}^{(2b)}$
\\
\hline
$\mathcal X_{1}^{(2b)}$
& $0$
& $0$
& $0$
& $\mathcal X_{1}^{(2b)}$
\\
$\mathcal X_{2}^{(2b)}$
& $0$
& $0$
& $(1-\nu)\mathcal X_{3}^{(2b)}$
& $0$
\\
$\mathcal X_{3}^{(2b)}$
& $0$
& $-(1-\nu)\mathcal X_{3}^{(2b)}$
& $0$
& $0$
\\
$\mathcal X_{4}^{(2b)}$
& $-\mathcal X_{1}^{(2b)}$
& $0$
& $0$
& $0$
\\
\hline
\end{tabular}
\end{table}
\end{case}
For the algebra \(\mathfrak g^{(2a)}\), the nonzero commutators are
\[
[\mathcal X_{1}^{(2a)},\mathcal X_{4}^{(2a)}]
=
B_0\mathcal X_{4}^{(2a)},
\qquad
[\mathcal X_{2}^{(2a)},\mathcal X_{3}^{(2a)}]
=
(1-\nu)\mathcal X_{3}^{(2a)}.
\]
Thus \(\mathfrak g^{(2a)}\) is the direct sum of two two-dimensional
solvable algebras. Using the adjoint action, a one-dimensional optimal
system is
\[
\left\{
\mathcal X_{1}^{(2a)}+\alpha\mathcal X_{2}^{(2a)},
\mathcal X_{1}^{(2a)}+\mathcal X_{3}^{(2a)},
\mathcal X_{2}^{(2a)}+\mathcal X_{4}^{(2a)},\mathcal X_{2}^{(2a)},
\mathcal X_{3}^{(2a)},
\mathcal X_{4}^{(2a)},
\mathcal X_{3}^{(2a)}+\mathcal X_{4}^{(2a)}
\right\},
\;
\alpha\in\mathbb R .
\]
For the algebra \(\mathfrak g^{(2b)}\), the nonzero commutators are
\[
[\mathcal X_{1}^{(2b)},\mathcal X_{4}^{(2b)}]
=
\mathcal X_{1}^{(2b)},
\qquad
[\mathcal X_{2}^{(2b)},\mathcal X_{3}^{(2b)}]
=
(1-\nu)\mathcal X_{3}^{(2b)}.
\]
Using the corresponding adjoint action, a one-dimensional optimal system is
\[
\left\{
\mathcal X_{4}^{(2b)}+\alpha\mathcal X_{2}^{(2b)},
\mathcal X_{4}^{(2b)}+\mathcal X_{3}^{(2b)},
\mathcal X_{2}^{(2b)}+\mathcal X_{1}^{(2b)},
\mathcal X_{1}^{(2b)},\mathcal X_{2}^{(2b)},
\mathcal X_{3}^{(2b)},
\mathcal X_{1}^{(2b)}+\mathcal X_{3}^{(2b)}
\right\},
\;
\alpha\in\mathbb R .
\]
\begin{case}\normalfont If $b=\nu,\; a=0,$ then
\begin{align}
   \xi=c_1z+c_2z\ln z,\\
   \eta=
\left(
\tau_t
-2c_1
-2c_2(\ln z+1)
\right)G(u)
\end{align}
Substituting these into \eqref{rad:5}, we obtain
\begin{align}
    \label{rad:15} AG\tau_{tt}
+\frac{2c_2(\nu-1)G}{z^2}
+\Bigl[\tau_t-2c_1-2c_2(\ln z+1)\Bigr]
(BG'+B-GB')
-\tau_tB=0.
\end{align}
Separation with respect to the independent functions
\(z^{-2}\), \(\ln z\), and \(1\) yields
\[
\nu=1,
\qquad
BG'+B-GB'=0.
\]
The latter equation integrates to
\begin{align*}
    B=B_0AG,\;R=B_0CG.
\end{align*}
Consequently, \eqref{rad:5} reduces to
\[
\tau_{tt}-B_0\tau_t=0,
\]
whose solution is
\[
\tau=
\begin{cases}
c_0+c_3e^{B_0t},
&
B_0\neq0,
\\[1mm]
c_0+c_3t,
&
B_0=0.
\end{cases}
\]
Furthermore, combining \eqref{rad:4} and \eqref{rad:9}, 
we obtain
\[
A(u)=A_0e^{u/G_0}.
\]
Therefore
\[
C(u)=A(u)K(u)=A_0K_0e^{u/G_0}.
\]
Thus, in the logarithmic branch,
\[
G=G_0,\qquad
K=K_0,\qquad
C=A_0K_0e^{u/G_0}.
\]
Finally,
\begin{align}
\xi&=
c_1z+c_2z\ln z,
\\[1mm]
\tau&=
\begin{cases}
c_0+c_3e^{B_0t},
&
B_0\neq0,
\\[1mm]
c_0+c_3t,
&
B_0=0,
\end{cases}
\\[1mm]
\eta&=
\begin{cases}
\Bigl(
B_0c_3e^{B_0t}
-2c_1
-2c_2(\ln z+1)
\Bigr)G_{0},
&
B_0\neq0,
\\[1mm]
\Bigl(
c_3
-2c_1
-2c_2(\ln z+1)
\Bigr)G_{0},
&
B_0=0.
\end{cases}
\end{align}
\end{case}
\noindent For \(B_0\neq0\), the admitted Lie algebra
\[
\mathfrak{g}^{(3a)}
=
\left\langle
\mathcal{X}_{1}^{(3a)},
\mathcal{X}_{2}^{(3a)},
\mathcal{X}_{3}^{(3a)},
\mathcal{X}_{4}^{(3a)}
\right\rangle
\]
is generated by
\begin{align}
\begin{cases}
\mathcal{X}_{1}^{(3a)}
=
\partial_t,
\\[2mm]
\mathcal{X}_{2}^{(3a)}
=
z\partial_z
-2G_0\partial_u,
\\[2mm]
\mathcal{X}_{3}^{(3a)}
=
z\ln z\,\partial_z
-2G_0(\ln z+1)\partial_u,
\\[2mm]
\mathcal{X}_{4}^{(3a)}
=
e^{B_0t}\partial_t
+
B_0G_0e^{B_0t}\partial_u.
\end{cases}
\label{alg:3a}
\end{align}

\noindent For \(B_0=0\), the admitted Lie algebra
\[
\mathfrak{g}^{(3b)}
=
\left\langle
\mathcal{X}_{1}^{(3b)},
\mathcal{X}_{2}^{(3b)},
\mathcal{X}_{3}^{(3b)},
\mathcal{X}_{4}^{(3b)}
\right\rangle
\]
is generated by
\begin{align}
\begin{cases}
\mathcal{X}_{1}^{(3b)}
=
\partial_t,
\\[2mm]
\mathcal{X}_{2}^{(3b)}
=
z\partial_z
-2G_0\partial_u,
\\[2mm]
\mathcal{X}_{3}^{(3b)}
=
z\ln z\,\partial_z
-2G_0(\ln z+1)\partial_u,
\\[2mm]
\mathcal{X}_{4}^{(3b)}
=
t\partial_t
+
G_0\partial_u.
\end{cases}
\label{alg:3b}
\end{align}
The Lie commutation relations of the admitted Lie algebras
$\mathfrak{g}^{(3a)}$ and $\mathfrak{g}^{(3b)}$
are summarized in Tables~\ref{tab:comm_3a}
and~\ref{tab:comm_3b}, respectively.
\begin{table}[htbp]
\centering
\caption{Commutator table for the Lie algebra
$\mathfrak{g}^{(3a)}
=
\langle
\mathcal X_{1}^{(3a)},
\mathcal X_{2}^{(3a)},
\mathcal X_{3}^{(3a)},
\mathcal X_{4}^{(3a)}
\rangle$.}
\label{tab:comm_3a}
\renewcommand{\arraystretch}{1.2}
\begin{tabular}{c|cccc}
\hline
$[\cdot,\cdot]$
& $\mathcal X_{1}^{(3a)}$
& $\mathcal X_{2}^{(3a)}$
& $\mathcal X_{3}^{(3a)}$
& $\mathcal X_{4}^{(3a)}$
\\
\hline
$\mathcal X_{1}^{(3a)}$
& $0$
& $0$
& $0$
& $B_0\mathcal X_{4}^{(3a)}$
\\
$\mathcal X_{2}^{(3a)}$
& $0$
& $0$
& $\mathcal X_{2}^{(3a)}$
& $0$
\\
$\mathcal X_{3}^{(3a)}$
& $0$
& $-\mathcal X_{2}^{(3a)}$
& $0$
& $0$
\\
$\mathcal X_{4}^{(3a)}$
& $-B_0\mathcal X_{4}^{(3a)}$
& $0$
& $0$
& $0$
\\
\hline
\end{tabular}
\end{table}
\begin{table}[htbp]
\centering
\caption{Commutator table for the Lie algebra
$\mathfrak{g}^{(3b)}
=
\langle
\mathcal X_{1}^{(3b)},
\mathcal X_{2}^{(3b)},
\mathcal X_{3}^{(3b)},
\mathcal X_{4}^{(3b)}
\rangle$.}
\label{tab:comm_3b}
\renewcommand{\arraystretch}{1.2}
\begin{tabular}{c|cccc}
\hline
$[\cdot,\cdot]$
& $\mathcal X_{1}^{(3b)}$
& $\mathcal X_{2}^{(3b)}$
& $\mathcal X_{3}^{(3b)}$
& $\mathcal X_{4}^{(3b)}$
\\
\hline
$\mathcal X_{1}^{(3b)}$
& $0$
& $0$
& $0$
& $\mathcal X_{1}^{(3b)}$
\\
$\mathcal X_{2}^{(3b)}$
& $0$
& $0$
& $\mathcal X_{2}^{(3b)}$
& $0$
\\
$\mathcal X_{3}^{(3b)}$
& $0$
& $-\mathcal X_{2}^{(3b)}$
& $0$
& $0$
\\
$\mathcal X_{4}^{(3b)}$
& $-\mathcal X_{1}^{(3b)}$
& $0$
& $0$
& $0$
\\
\hline
\end{tabular}
\end{table}
The commutation relations listed in Tables~\ref{tab:comm_3a}
and~\ref{tab:comm_3b} show that both Lie algebras
\(\mathfrak g^{(3a)}\) and \(\mathfrak g^{(3b)}\) are
decomposable and isomorphic to the direct sum
\(A_{2.1}\oplus A_{2.1}\) of two non-Abelian
two-dimensional solvable Lie algebras \cite{mubarakzyanov1963solvable,patera1975continuous,patera1975continuous2,patera1977continuous,patera1977subalgebras}. Therefore,
the classification of one-dimensional subalgebras can
be carried out by means of the adjoint representation.
Following the standard procedure of Olver \cite{olverbook},
equivalent subalgebras are identified under the action
of the corresponding inner automorphism group, leading
to the optimal systems presented in \autoref{opt_sys:3a+3b}.
\begin{table}[H]
\centering
\caption{Optimal systems of one-dimensional subalgebras.}
\begin{tabular}{ll}
\hline
Case (3a) & Case (3b)\\
\hline
$\mathcal X_{1}^{(3a)}$ &
$\mathcal X_{1}^{(3b)}$\\
$\mathcal X_{2}^{(3a)}$ &
$\mathcal X_{2}^{(3b)}$\\
$\mathcal X_{3}^{(3a)}$ &
$\mathcal X_{3}^{(3b)}$\\
$\mathcal X_{4}^{(3a)}$ &
$\mathcal X_{4}^{(3b)}$\\
$\mathcal X_{1}^{(3a)}+\mathcal X_{2}^{(3a)}$ &
$\mathcal X_{1}^{(3b)}+\mathcal X_{2}^{(3b)}$\\
$\mathcal X_{1}^{(3a)}+\mathcal X_{3}^{(3a)}$ &
$\mathcal X_{1}^{(3b)}+\mathcal X_{3}^{(3b)}$\\
$\mathcal X_{2}^{(3a)}+\mathcal X_{4}^{(3a)}$ &
$\mathcal X_{2}^{(3b)}+\mathcal X_{4}^{(3b)}$\\
$\mathcal X_{3}^{(3a)}+\mathcal X_{4}^{(3a)}$ &
$\mathcal X_{3}^{(3b)}+\mathcal X_{4}^{(3b)}$\\
\hline
\label{opt_sys:3a+3b}
\end{tabular}
\end{table}
The optimal systems for Case (3a) and Case (3b) coincide because both
\(\mathfrak g^{(3a)}\) and \(\mathfrak g^{(3b)}\) are isomorphic to
\(A_{2.1}\oplus A_{2.1}\).
\section{Similarity Reductions and Invariant Solutions}
After identifying the admitted Lie algebras and their associated optimal systems of one-dimensional subalgebras, the subsequent step is to develop group-invariant solutions of the governing equation \eqref{Rad:01}. For each generator in the optimal system, the corresponding invariants are derived from the characteristic equations:
\begin{align}
    \label{rad:17}\frac{dt}{\tau}
=
\frac{dz}{\xi}
=
\frac{du}{\eta}.
\end{align}
These invariants produce similarity variables and associated similarity representations of the dependent variable. The substitution of the resultant ans\"atze into the governing equation transforms the original partial differential equation into an ordinary differential equation, whose solutions yield precise group-invariant solutions. As every one-dimensional subalgebra corresponds to a representative element of the optimal system, it is adequate to do the reduction solely for the generators included within it. Before starting similarity reduction we need to divide \eqref{eq:governing_expanded} both sides by $K(u)$ and it becomes:
\begin{align}
    \label{rad:16} A(u)u_t
=
u_{zz}
+\frac{K'(u)}{K(u)}\,u_z^2
+\frac{\nu}{z}\,u_z
+B_0A(u)G(u),
\end{align}
this is done to obtain a pure ODE, because the functions 
$A(u), K(u), G(u)$ must be replaced by their invariant functional forms corresponding to the symmetry. In the following, we present similarity reductions and group-invariant solutions associated with selected symmetry generators.
\subsection{Case I:
$\mathcal{X}^{(1)}_{1}+\alpha \mathcal{X}^{(1)}_{2}$}
\noindent We consider the one-parameter symmetry group formed by $\mathcal{X}^{(1)}_{1}+\alpha \mathcal{X}^{(1)}_{2}$ and derive the associated similarity variables and group-invariant solutions.
\begin{align}\label{rad:22}
    \mathcal{X}=\,\mathcal{X}^{(1)}_{1}+\alpha \mathcal{X}^{(1)}_{2}=\, \partial_t
+\alpha z\,\partial_z
-\alpha\left(\frac{b-1}{2}\,u+2G_0\right)\partial_u.
\end{align}
The characteristics system \eqref{rad:17} becomes
\begin{align}
\label{rad:18}\frac{dt}{1}
=
\frac{dz}{\alpha z}
=
\frac{du}{-\alpha\left(mu+2G_0\right)}, \quad m=\frac{b-1}{2}.
\end{align}
The corresponding invariant form of $u$ and the similarity variable can be obtained as follows:
\begin{align}
\label{rad:19}u(z,t)
=
e^{-qt}\phi(\omega)
-\frac{4G_0}{b-1},
\quad
\omega
=
z e^{-\alpha t},\qquad q=\,\frac{\alpha\,(b-1)}{2}.
\end{align}
At this stage, we can not  substitute the similarity ans\"atze \eqref{rad:19} directly into PDE \eqref{rad:16} while leaving $A(u), K(u), G(u)$ arbitrary, instead we first need to find admissible forms of these functions. Using the notation introduced in Eqs.~\eqref{rad:20} and \eqref{rad:21}, we have
\begin{align*}
    F(u)=\frac{C'(u)}{C(u)}-\frac{K'(u)}{K(u)}=\frac{A'(u)}{A(u)}, \quad \text{and}\,G(u)=\frac1{F(u)}
\end{align*}
To reduce \eqref{rad:16} into pure ODE we must find the invariant form of $A(u)$ under symmetry generator \eqref{rad:22}. Due to the invariance of PDE \eqref{rad:16} the coefficient $A(u)$ must transform consistently under the prolonged action of symmetry generator \eqref{rad:22}, and we find that
\begin{align}
    \label{rad:23}
\operatorname{pr}^{(1)}\mathcal{X}\!\left(Au_t\right)
=
\left[
-\alpha\,\left(mu+2G_0\right)A'(u)
-\alpha mA(u)
\right]u_t.
\end{align}
 and 
\begin{align}
    \label{rad:24}
\operatorname{pr}^{(2)}\mathcal{X}\!\left(u_{zz}\right)
=-\alpha\,(m+2)\,u_{zz}.
\end{align}
For invariance, the term $A(u)\,u_{t}$ must scale with same weight as $u_{zz}$, therefore, the equations \eqref{rad:23} and \eqref{rad:24} yields
\begin{align}
    \label{rad:25} \frac{-\alpha\left[(m\,u+2G_0)A'+m\right]}{A}
=
-\alpha(m+2)
\end{align}
The solution of \eqref{rad:25} will provided he homogeneous transformation requirement for $A(u)$
\begin{align}
    \label{rad:28}A(u)={A}_0\left(u+\frac{4G_0}{b-1}\right)^{2}.
\end{align}
Define shift variable
\begin{align}
    \label{rad:26}v=u+\frac{4G_0}{b-1}.
\end{align}
So that the similarity ans\"atze \eqref{rad:19} becomes:
\begin{align}
    \label{rad:27}v=e^{-\alpha m t}\phi(\omega),
\qquad
\omega=ze^{-\alpha t}.
\end{align}
The coefficient \eqref{rad:28} reduce to
\begin{align}
     A(u)=A_0 e^{-\alpha (b-1)t}\phi(\omega)^2 \nonumber\\
   \label{rad:30} \implies A(u)v_t
=
- A_0 \alpha e^{-3\alpha m t}
\phi^2
\left(
m\phi+\omega\phi'
\right).
\end{align}
whereas the diffusion term $v_{zz}$ scale as 
\begin{align}
    e^{-\alpha(m+2)t}
\end{align}
this gives
\begin{align*}
    3\,m=\,m+2 \implies m=1 \implies b=3.
\end{align*}
This is the crucial compatibility condition for reduction into a pure ODE  under  symmetry generator \eqref{rad:22}. Now we express $\frac{K'}{K}$ in invariant form. Let
\[
P(u)=\frac{K'(u)}{K(u)}.
\]
Since \(C=AK\), we have
\[
\frac{C'}{C}=\frac{A'}{A}+\frac{K'}{K}.
\]
Hence
\[
\frac{K''}{K}-\frac{C'}{C}\frac{K'}{K}
=
P'+P^2-\left(\frac{A'}{A}+P\right)P
=
P'-\frac{A'}{A}P.
\]
Therefore, the determining equation \eqref{rad:4}
\[
G''+\frac{K'}{K}G'
+\left(
\frac{K''}{K}
-\frac{C'}{C}\frac{K'}{K}
\right)G=0
\]
reduces to
\[
G''+PG'
+\left(P'-\frac{A'}{A}P\right)G=0.
\]
For the present invariant class, \(b=3\). Thus
\[
v=u+2G_0,\qquad A=A_0v^2.
\]
Therefore
\[
\frac{A'}{A}=\frac{2}{v},\qquad
G=\frac{v}{2},\qquad
G'=\frac{1}{2},\qquad G''=0.
\]
Substituting these into the reduced determining equation gives
\[
\frac{P}{2}
+
\left(P'-\frac{2}{v}P\right)\frac{v}{2}
=0.
\]
Thus
\[
\frac{vP'}{2}-\frac{P}{2}=0,
\]
or
\[
vP'-P=0.
\]
Hence
\[
\frac{P'}{P}=\frac{1}{v},
\]
which gives
\[
P=\lambda v.
\]
Consequently,
\[
\frac{K'}{K}
=
\lambda\left(u+2G_0\right),
\qquad b=3,
\]
where \(\lambda\) is an arbitrary constant. Consequently, upon integration
\begin{align}
    &K(u)
=
K_0
\exp\!\left[
\frac{\lambda}{2}\left(u+2G_0\right)^2
\right],\\
&C(u)
=
A_0K_0\left(u+2G_0\right)^2
\exp\!\left[
\frac{\lambda}{2}\left(u+2G_0\right)^2
\right].
\end{align}
But for the term $\frac{K'}{K}u_{z}^{2}$ to scale like the diffusion term we need to set $\lambda=0$. Finally, substituting the similarity ans\"atze \eqref{rad:19} into the governing equation \eqref{rad:16}
and collecting the resulting terms, we obtain the reduced equation
\begin{align}
   \label{rad:31} \phi''+\frac{\nu}{\omega}\phi'
+\alpha A_0\omega\phi^2\phi'
+A_0\left(\alpha+\frac{B_0}{2}\right)\phi^3=0.
\end{align}
with the admissible coefficient functions
\begin{align*}
    &A(u)=A_0\left(u+2G_0\right)^2,\\
   & K(u)=K_0, C(u)=C_0\left(u+2G_0\right)^2,\\
    &R(u)=\frac{B_0C_0}{2}\left(u+2G_0\right)^3.
\end{align*}
The reduced equation \eqref{rad:31} is a nonlinear second-order Emden-Fowler-type equation, more precisely a generalized Lane-Emden equation for $\beta=\,\alpha\,A_{0}$ and $\gamma=\,A_{0}\left(\alpha+\frac{B_{0}}{2}\right)$ \cite{zaitsev2002handbook} with an additional convective nonlinear term $\omega\phi^2\phi'$. 

\noindent Multiply the equation \eqref{rad:31} by \(\omega^{\nu}\):
\[
\omega^{\nu}\phi''
+\nu\omega^{\nu-1}\phi'
+\beta\omega^{\nu+1}\phi^{2}\phi'
+\gamma\omega^{\nu}\phi^{3}
=0.
\]
The first two terms combine according to
\[
(\omega^{\nu}\phi')'
=
\omega^{\nu}\phi''
+\nu\omega^{\nu-1}\phi'.
\]

\noindent Hence
\[
(\omega^{\nu}\phi')'
+\beta\omega^{\nu+1}\phi^{2}\phi'
+\gamma\omega^{\nu}\phi^{3}
=0.
\]

\noindent Now observe that
\[
\phi^{2}\phi'
=
\frac{1}{3}(\phi^{3})'.
\]

\noindent Therefore,
\[
\beta\omega^{\nu+1}\phi^{2}\phi'
=
\frac{\beta}{3}\omega^{\nu+1}(\phi^{3})'.
\]

\noindent Using the product rule,
\[
\omega^{\nu+1}(\phi^{3})'
=
(\omega^{\nu+1}\phi^{3})'
-(\nu+1)\omega^{\nu}\phi^{3}.
\]
Substituting this identity produces
\begin{align}
  \label{rad:34}(\omega^{\nu}\phi')'
+\frac{\beta}{3}(\omega^{\nu+1}\phi^{3})'
+\left[
\gamma-\frac{\beta(\nu+1)}{3}
\right]
\omega^{\nu}\phi^{3}
=0.  
\end{align}
This equation is exactly solvable if $\gamma=\frac{\beta(\nu+1)}{3}$. It can be solved easily by direct integration and solution is obtained as follows:
\begin{align}
    \label{rad:33}\phi(\omega)=\pm\frac{1}{\sqrt{C_3+\frac{\beta}{3}\omega^2}}
\end{align}
In original notations the compatibility condition must be 
\begin{align*}
    A_0\left(\alpha+\frac{B_0}{2}\right)=
\frac{\alpha A_0(\nu+1)}{3}\implies B_{0}=\,\frac{2\,\alpha(\nu-2)}{3}
\end{align*}
The original solution is thus can be written as:
\begin{align}
    u(z,t)
=
-2G_0
\pm
\frac{1}
{\sqrt{
C_3 e^{2\alpha t}
+\frac{\alpha A_0}{3}z^2
}}.
\end{align}
This is a rich family of solutions for the governing equation \eqref{eq:governing_expanded} and it is physically relevant with regular bounded profile \cite{barenblatt1996scaling,vazquez2006porous}.
\subsection{Case II: $\mathcal{X}^{(1)}_{3}+\beta \mathcal{X}^{(1)}_{2}$}
\noindent The symmetry reduction for the generator \(\mathcal{X}^{(1)}_{3}+\beta \mathcal{X}^{(1)}_{2}\) works similarly to the preceding case. We skip the intermediate calculations and show simply the similarity variables, the reduced ordinary differential equation, and the group-invariant solutions are presented below in this section. The symmetry generator in this case can be written as
\begin{align}\label{sec_similarity}
\mathcal{X}
&=
e^{B_0 t}\partial_t
+\beta z\,\partial_z
+\left(B_0 e^{B_0 t}-2\beta\right)\Theta(u)\,\partial_u,\quad \Theta(u)=\frac{b-1}{4}\,u+G_0,.
\end{align}
The similarity transformations associated with this symmetry generator are obtained by solving the characteristic equations \eqref{rad:17}, resulting in the following.
\begin{equation}\label{rad:37}
    \begin{aligned}
u(z,t)
&=
e^{\frac{(b-1)B_0}{4}t}
\exp\!\left(
\frac{(b-1)\beta}{2B_0}e^{-B_0 t}
\right)
\phi(\omega)
-\frac{4G_0}{b-1},
\\[2mm]
\omega
&=
z\exp\!\left(
\frac{\beta}{B_0}e^{-B_0 t}
\right).
\end{aligned}
\end{equation}
and the corresponding reduced equation 
\begin{align}
\label{rad:35}\phi''
+\frac{\nu}{\omega}\phi'
+\beta A_0\,\phi^{\frac{4}{b-1}}
\left(
\omega\phi'
+\frac{b-1}{2}\phi
\right)
=0.
\end{align}
The reduced equation \eqref{rad:35} is again nonlinear Emden-Fowler/Lane-Emden type ODE with a nonlinear convective term having more  general power-law than \eqref{rad:31}. It can be solved by first making some parameteric changes, we consider $n=\frac{4}{b-1}$. Then the reduced ODE
\[
\phi''
+\frac{\nu}{\omega}\phi'
+\beta A_0\phi^{n}
\left(
\omega\phi'
+\frac{b-1}{2}\phi
\right)=0
\]
can be rewritten in divergence form as
\[
\left(\omega^\nu \phi'\right)'
+\frac{\beta A_0}{n+1}
\left(\omega^{\nu+1}\phi^{\,n+1}\right)'
=0,
\]
provided that
\[
\frac{b-1}{2}
=
\frac{\nu+1}{n+1}.
\]
Substituting \(n=4/(b-1)\), this condition reduces to
\[
b=2\nu-1.
\]
Integrating once yields
\[
\omega^\nu \phi'
+\frac{\beta A_0}{n+1}\,
\omega^{\nu+1}\phi^{\,n+1}
=C_1,
\]
where \(C_1\) is an integration constant. For the particular case \(C_1=0\), we obtain the separable equation
\[
\phi'
=
-\frac{\beta A_0}{n+1}\,
\omega\phi^{\,n+1}.
\]
Integrating gives
\[
\phi^{-n}
=
C_2
+\frac{n\beta A_0}{2(n+1)}\,\omega^2,
\]
or equivalently
\[
\phi(\omega)
=
\left(
C_2
+\frac{n\beta A_0}{2(n+1)}\,\omega^2
\right)^{-1/n}.
\]
Substituting \(n=4/(b-1)\), we arrive at
\begin{align}\label{rad:36}
\phi(\omega)
=
\left(
C_2
+\frac{2\beta A_0}{b+3}\,\omega^2
\right)^{-\frac{b-1}{4}},
\qquad
b=2\nu-1.
\end{align}
The original solution of PDE \eqref{rad:16} can be obtained by combining \eqref{rad:37} with \eqref{rad:36}.
\subsection{Case III: $\mathcal X_{2}^{(2a)}+\mathcal X_{4}^{(2a)}
$} \noindent The reduction corresponding to this symmetry belongs to the branch 

\begin{align}
    a=0,\; b=\,\frac{\nu}{2-\nu},
\end{align}
and for a valid symmetry reduction, we need to take $K'(u)=0$. In the subcase \(a=0\), one has
\[
K(u)=K_0,
\qquad
C(u)=K_0A(u).
\]
Moreover, since
\[
\frac{A'}{A}=\frac1G,
\]
we obtain
\[
A(u)=A_0G(u)^{1/\alpha},
\qquad
\alpha=\frac{b-1}{4}.
\]
The symmetry generator under consideration is
\begin{align}
\mathcal{X}
=
\mathcal{X}_{2}^{(2a)}
+
\mathcal{X}_{4}^{(2a)}
=
e^{B_0t}\partial_t
+
z\partial_z
+
\left(B_0e^{B_0t}-2\right)H(u)\partial_u,
\end{align}
where
\[
H(u)=\frac{b-1}{4}u+G_0=\alpha u+G_0.
\]
Solving the characteristic equations \eqref{rad:17} gives the and the invariant form and similarity variable
\begin{align}
\label{rad:38}H(u)
=
\exp\left(
\alpha B_0t
+
\frac{2\alpha}{B_0}e^{-B_0t}
\right)\phi(\omega),\quad \omega
=
\ln z+\frac{e^{-B_0t}}{B_0},
\end{align}

Using this ans\"atze  \eqref{rad:38} in \eqref{rad:16} gives reduces ODE as follows:

\begin{align}
\label{rad:39}\phi''
+
(\nu-1)\phi'
+
A_0e^{2\omega}
\phi^{\frac{2(2-\nu)}{\nu-1}}
\left(
\phi'
+
\frac{\nu-1}{2-\nu}\phi
\right)
=0,
\qquad
\nu\neq1,2.
\end{align}
The similarity reduction obtained from the generator
\(\mathcal{X}_{2}^{(2a)}+\mathcal{X}_{4}^{(2a)}\) is valid only for
\(\nu\neq1,2\). From a physical perspective, these values correspond to important radial
geometries and are therefore expected to possess distinct symmetry
structures. As far as solution of equation \eqref{rad:39} is concerned, it is a nonlinear non-autonomous second-order ordinary
differential equation containing derivative-dependent power-law
nonlinearities. The presence of the explicit factor
\(e^{2\omega}\), the mixed nonlinear term \(\phi^{m}\phi'\), and the
parameter-dependent exponent \(m\) prevents the reduction to any standard
integrable class. As a result of this, the closed-form solutions
can be expected only for special parameter values or under additional
ans\"atze. 

The  equation \eqref{rad:39} can be transformed to simpler form by making substitution:
\begin{align}
    \label{rad:42}r=\,\frac{\nu-1}{2-\nu}
\end{align}
and it reduced to
\begin{align}
    \label{rad:41} \phi''+(\nu-1)\phi'+A_0e^{2\omega}\phi^{2/r}\left(\phi'+r\phi\right)
=0.
\end{align}
by making dependent variable transformation;
\begin{align*}
    \phi=e^{-r\omega}y
\end{align*}
it further reduce to
\begin{align}
    y''+(\nu-1-2r)y'+\left(r^2-(\nu-1)r\right)y+A_0y^{2/r}y'=0.
\end{align}
Invoking the value of $r$ from \eqref{rad:42}, the most simpler version of \eqref{rad:39} can be obtained as follows:
\begin{align}
    \label{rad:43}y''-\frac{\nu(\nu-1)}{2-\nu}\,y'+\frac{(\nu-1)^3}{(2-\nu)^2}\,y+A_0y^{\frac{2(2-\nu)}{\nu-1}}y'=0.
\end{align}
It is equivalent to following autonomous Lienard-type equation \cite{zaitsev2002handbook}
\begin{align*}
    y''+f(y)y'+g(y)=0,
\end{align*}
where 
\begin{align*}
    f(y)=\,-\frac{\nu(\nu-1)}{2-\nu}+A_0y^{\frac{2(2-\nu)}{\nu-1}},\;g(y)=\,\frac{(\nu-1)^3}{(2-\nu)^2}
\end{align*}
For specific values of the geometric parameter $\nu$, the reduced equation \eqref{rad:43} significantly simplifies, resulting in numerous notable subclasses of nonlinear ordinary differential equations.
\begin{align}
    \label{rad:44}y''+A_{0}\,\frac{1}{y^{4}}\,y'-\frac{1}{8}\,y=0,
\end{align}
which is a Lienard equation with inverse-power damping. Second useful subclass can be obtained when power of $y$ in \eqref{rad:43} becomes unity, that is when $\nu=\frac{5}{3}$, we get 
\begin{align}
    \label{rad:45}y''+(A_{0}\,y-10)\,y'+8\,y=0.
\end{align}
The third important subclass can obtained, when power of $y$ becomes 2, that is when $\nu=\frac{3}{2}$, we get
\begin{align}
    \label{rad:46}y''+(A_{0}\,y^{2}-3)\,y'+\frac{1}{2}\,y=0,
\end{align}
which resembles with generalized Rayleigh/Lienard oscillator.
\subsection{Case IV: $\mathcal X_{3}^{(3a)}+\mathcal X_{4}^{(3a)}
$ } \noindent This reduction corresponds to the branch $\nu=1$ with the admissible coefficient functions
\begin{align*}
    &G(u)=G_0,\qquad K(u)=K_0,\\
    &A(u)=A_0e^{u/G_0},
C(u)=A_0K_0e^{u/G_0},
R(u)=B_0C(u)G_0.
\end{align*}
Consider the generator

\begin{align}
    \mathcal X= \mathcal X_{3}^{(3a)}+\mathcal X_{4}^{(3a)}
=
e^{B_0t}\partial_t
+
z\ln z\,\partial_z
+
\left[
B_0G_0e^{B_0t}
-2G_0(\ln z+1)
\right]\partial_u
\end{align}
The invariant form of $u$ and the similarity variable is thus obtained as follows:
\begin{align}
    \label{rad:40} u(z,t)
=
\phi(\omega)
+
G_0\left(
B_0t-2\ln z-2\ln(\ln z)
\right),
\quad
\omega=
\ln(\ln z)+\frac{e^{-B_0t}}{B_0}
\end{align}
Substituting the similarity ans\"atz \eqref{rad:40} into PDE \eqref{rad:16} produces the reduced ODE
\begin{align}
   \phi''+\left(A_0e^{\phi/G_0}-1\right)\phi'+2G_0=0.
\end{align}
This equation can be transformed into a simpler form through the substitution $\phi = G_{0}\ln y$. However, even after this transformation, obtaining a closed-form solution remains unlikely unless additional restrictions or suitable parametric assumptions are imposed.

\section{Observation and analysis}
Compared to the conventional symmetry analysis of equations with fixed coefficients, the Lie symmetry classification taken into consideration in this work is significantly more complex. The main challenge is from the existence of arbitrary constitutive functions $C(u), K(u)$, and $R(u)$. These functions transforms the determining equations into a coupled functional-differential system. It is typically hard to identify the symmetry generators without concurrently placing limits on the coefficient functions since the infinitesimals are closely associated with the constitutive ratios $A=\,CK^{-1}$ and $B=\,RK^{-1}$.

A subsequent difficulty emerges from the radial geometry term $\nu z^{-1}u_z$, which destroys translational invariance and generates specific geometric branches associated with particular values of $\nu$. Consequently, the generic classification separates naturally into
several exceptional subclasses, such as the logarithmic branch for
\(\nu=1\) and the singular branch associated with \(\nu=2\).

Although the introduction of auxiliary variables such as
\[
A(u)=\frac{C(u)}{K(u)},
\qquad
B(u)=\frac{R(u)}{K(u)},
\qquad
G(u)=\left(\frac{C'}{C}-\frac{K'}{K}\right)^{-1}
\]
substantially simplified the determining equations; yet, the resultant system continues to be extremely nonlinear and strongly coupled. Consequently, explicit classification is feasible only upon the imposition of additional compatibility constraints or the consideration of symmetry-enhancing subclasses.

A further limitation of the approach is that the reduced ordinary differential equations such as \eqref{rad:39} derived from similarity reductions are, generally, nonlinear non-autonomous equations characterized by derivative-dependent power-law nonlinearities. Since these equations do not fall into the conventional integrable class, exact closed-form invariant solutions are only possible with specific parameter choices or additional assumptions.
Despite difficulties, the procedure successfully identifies nontrivial symmetry extensions, admissible constitutive relations, and inequivalent similarity reductions, providing a systematic framework for analyzing generalized radial heat equations with nonlinear reaction sources.
\section{Conclusion}
In this paper, Lie symmetry analysis was applied to a generalized nonlinear radial heat equation with an additional nonlinear reaction source. The presence of the arbitrary functions $C(u), K(u)$, and $R(u)$ led to a coupled functional-differential classification problem. By introducing reduced constitutive variables, the determining system was simplified and several symmetry-enhancing subclasses were obtained.

The analysis showed that the radial geometry parameter plays an important role in the classification. In particular, special branches arise for exceptional values of the parameter, including logarithmic and singular cases. For each admissible subclass, the corresponding Lie symmetry generators were derived, their commutation relations were studied, and optimal systems of one-dimensional subalgebras were constructed. These optimal systems were then used to obtain similarity variables and reduce the governing equation to ordinary differential equations.

The similarity reductions produced nonlinear Emden-Fowler, Lane-Emden-type, and Lienard-type  equations. For selected parameter restrictions, exact invariant solutions were obtained, illustrating the usefulness of the symmetry approach. Overall, the results demonstrate that the nonlinear source term introduces new compatibility conditions and generates richer symmetry structures than those found in the source-free model. The classification presented here provides a systematic basis for constructing exact solutions and for further analytical study of nonlinear heat-transfer models in radial geometries.
\subsection*{Funding}Not applicable.
\subsection*{Availability of data and material} Not applicable.
\section*{Declaration}
\subsection*{Conflict of interest} The authors declare that there is no conflict of interest with respect to the publication of this manuscript.
\subsection*{Funding Statement}This research did not receive external funding.

 \bibliography{My.Bibtex.Library} 

@article{mansfield2,
author = {Clarkson, P. and Mansfield, E.},
title = {Algorithms for the Nonclassical Method of Symmetry Reductions},
journal = {SIAM Journal on Applied Mathematics},
volume = {54},
number  = {6},
pages = {1693-1719},
year = {1994},
}

@book{haydonbook,
author = "P. E. Hydon",
title = {Symmetry Methods for Differential Equations},
publisher = {Cambridge University Press},
year = {2000},
address = {Cambridge}
}

@book{anco,
author = {Bluman, G.W. and  Anco, S C },
title ={\href{http://www.springer.com/us/book/9780387986548} {Symmetry and Integration Methods for Differential Equations}},
volume = {154},
publisher={Springer-Verlag Inc.},
year = {2002},
address={New York}
}

@article{NUCCI,
author = {Nucci, M.C. and   Clarkson, P.A. },
title ={\href{http://www.sciencedirect.com/science/article/pii/037596019290904Z} {The nonclassical method is more general than the direct method for
symmetry reductions: An example of the {Fitzhugh-Nagumo} equation}},
journal={Physics Letters A},
volume = {164},
number={1},
pages = {49-56},
year = {1992},
}

@book{olverbook,
author ={Olver, P.J.},
title = {\href{http://www.springer.com/us/book/9780387950006}{Applications of Lie Groups to Differential Equations}},
volume = {107},
publisher={Springer-Verlag Inc.},
year = {1986},
address={New York},
}

@book{ovsi,
author = {Ovsiannikov, L.V.},
title = {\href{http://store.elsevier.com/product.jsp?isbn=9781483219066&pagename=search}{Group Analysis of Differential Equations}},
publisher={Academic Press},
year = {1982},
address={New York},
}

@article{patera1977subalgebras,
  title={\href{http://scitation.aip.org/content/aip/journal/jmp/18/7/10.1063/1.523441}{Subalgebras of real three and four dimensional Lie algebras}},
  author={Patera, J and Winternitz, P},
  journal={Journal of Mathematical Physics},
  volume={18},
  number={7},
  pages={1449--1455},
  year={1977},
  publisher={AIP Publishing}
}

@article{bruzon2001symmetry,
  title={\href{http://iopscience.iop.org/article/10.1088/0305-4470/34/18/304/meta}{The symmetry reductions of a turbulence model}},
  author={Bruz{\'o}n, MS and Clarkson, PA and Gandarias, ML and Medina, E},
  journal={Journal of Physics A: Mathematical and General},
  volume={34},
  number={18},
  pages={3751-3760},
  year={2001},
  publisher={IOP Publishing}
}

@article{patera1975continuous,
  title={\href{http://scitation.aip.org/content/aip/journal/jmp/16/8/10.1063/1.522729}{Continuous subgroups of the fundamental groups of physics. I. General method and the Poincar{\'e} group}},
  author={Patera, J and Winternitz, Po and Zassenhaus, H},
  journal={Journal of Mathematical Physics},
  volume={16},
  number={8},
  pages={1597--1614},
  year={1975},
  publisher={AIP Publishing}
}

@article{patera1975continuous2,
  title={\href{http://scitation.aip.org/content/aip/journal/jmp/16/8/10.1063/1.522730}{Continuous subgroups of the fundamental groups of physics. II. The similitude group}},
  author={Patera, J and Winternitz, P and Zassenhaus, H},
  journal={Journal of Mathematical Physics},
  volume={16},
  number={8},
  pages={1615--1624},
  year={1975},
  publisher={AIP Publishing}
}

@article{patera1977continuous,
  title={\href{http://scitation.aip.org/content/aip/journal/jmp/18/12/10.1063/1.523237}{Continuous subgroups of the fundamental groups of physics. III. The de Sitter groups}},
  author={Patera, J and Sharp, RT and Winternitz, P and Zassenhaus, H},
  journal={Journal of Mathematical Physics},
  volume={18},
  number={12},
  pages={2259--2288},
  year={1977},
  publisher={AIP Publishing}
}

@article{bira2015exact,
  title={\href{http://link.springer.com/article/10.1007/s10483-015-1968-7}{Exact solutions to drift-flux multiphase flow models through Lie group symmetry analysis}},
  author={Bira, B and Sekhar, T Raja},
  journal={Applied Mathematics and Mechanics},
  volume={36},
  number={8},
  pages={1105--1112},
  year={2015},
  publisher={Springer}
}

@book{bluman2010applications,
  title={Applications of Symmetry Methods to Partial Differential Equations},
  author={Bluman, George and Cheviakov, Alexei F and Anco, Stephen C},
  volume={168},
  year={2010},
  publisher={Springer},
address={New York}
}

@article{ames1992symmetry,
  title={Symmetry in nonlinear mechanics},
  author={Ames, WF},
  journal={Mathematics in science and engineering},
  volume={185},
  pages={31--78},
  year={1992},
  publisher={Elsevier}
}

@article{pandey2008symmetry,
  title={Symmetry analysis and exact solutions of magnetogasdynamic equations},
  author={Pandey, Manoj and Radha, R and Sharma, V D},
  journal={The Quarterly Journal of Mechanics \& Applied Mathematics},
  volume={61},
  number={3},
  pages={291--310},
  year={2008},
  publisher={Oxford University Press}
}

@article{pandey2009symmetry,
  title={Symmetry groups and similarity solutions for the system of equations for a viscous compressible fluid},
  author={Pandey, Manoj and Pandey, BD and Sharma, V D},
  journal={Applied Mathematics and Computation},
  volume={215},
  number={2},
  pages={681--685},
  year={2009},
  publisher={Elsevier}
}

@article{faucher1993symmetry,
  title={Symmetry analysis of the {Infeld-Rowlands} equation},
  author={Faucher, M and Winternitz, P},
  journal={Physical Review E},
  volume={48},
  number={4},
  pages={3066-3071},
  year={1993},
  publisher={APS}
}

@article{mubarakzyanov1963solvable,
  title={On solvable Lie algebras},
  author={Mubarakzyanov, Gamir Mubarakzyanovich},
  journal={Izvestiya Vysshikh Uchebnykh Zavedenii. Matematika},
  number={1},
  pages={114--123},
  year={1963},
  publisher={Kazan (Volga region) Federal University}
}

@article{nauryz2026lie,
  title={Lie symmetry analysis of the nonlinear generalized heat equation for varying cross-section geometry},
  author={Nauryz, Targyn A},
  journal={arXiv preprint arXiv:2604.24418},
  year={2026}
}

@book{zaitsev2002handbook,
  title={Handbook of exact solutions for ordinary differential equations},
  author={Zaitsev, Valentin F and Polyanin, Andrei D},
  year={2002},
  publisher={Chapman and Hall/CRC}
}

@book{barenblatt1996scaling,
  title={Scaling, self-similarity, and intermediate asymptotics: dimensional analysis and intermediate asymptotics},
  author={Barenblatt, Grigory Isaakovich},
  number={14},
  year={1996},
  publisher={Cambridge University Press}
}

@book{vazquez2006porous,
  title={The porous medium equation: mathematical theory},
  author={V{\'a}zquez, Juan Luis},
  year={2006},
  publisher={Clarendon Press}
}
 \bibliographystyle{elsarticle-num}

\end{document}